\documentclass[11pt,a4paper,english,nofootinbib,,superscriptaddress]{revtex4}
\usepackage{lmodern}

\usepackage[T1]{fontenc}
\usepackage[latin9]{inputenc}
\usepackage{babel}

\usepackage{amsmath}
\usepackage{graphicx}
\usepackage{amssymb}
\usepackage{esint}
\usepackage[unicode=true, pdfusetitle,
 bookmarks=true,bookmarksnumbered=false,bookmarksopen=false,
 breaklinks=false,pdfborder={0 0 1},backref=false,colorlinks=false]
 {hyperref}
\usepackage{amsmath}
\usepackage{amssymb}
\def\b{\begin{equation}}
\def\e{\end{equation}}
\def\K{{\cal K}}

\def\K{{\cal K}}

\def\R{{\rm I\hspace{-.15em}R}}

\def\bd{\begin{displaystyle}}
\def\ed{\end{displaystyle}}
\def\ba{\begin{array}}
\def\ea{\end{array}}

\def\bee{\begin{enumerate}}
\def\eee{\end{enumerate}}

\def\bes{\begin{eqnarray*}}
\def\ees{\end{eqnarray*}}
\def\be{\begin{eqnarray}}
\def\ee{\end{eqnarray}}

\makeatletter
\@ifundefined{textcolor}{}
{%
 \definecolor{BLACK}{gray}{0}
 \definecolor{WHITE}{gray}{1}
 \definecolor{RED}{rgb}{1,0,0}
 \definecolor{GREEN}{rgb}{0,1,0}
 \definecolor{BLUE}{rgb}{0,0,1}
 \definecolor{CYAN}{cmyk}{1,0,0,0}
 \definecolor{MAGENTA}{cmyk}{0,1,0,0}
 \definecolor{YELLOW}{cmyk}{0,0,1,0}
 }

\usepackage{latexsym}\usepackage{bm}

\makeatother

\begin{document}

\title{Conformal linear gravity in de Sitter space II}

\author{M.V. Takook}

\email{takook@razi.ac.ir}

\affiliation{Department of Physics, Science and
Research Branch, Islamic Azad University, Tehran,
Iran}

\author{ H. Pejhan}

\affiliation{Department of Physics, Science and
Research Branch, Islamic Azad University, Tehran,
Iran}

\author{ M. Reza Tanhayi}

\email{ m_tanhayi@iauctb.ac.ir}

\affiliation{Department of Physics, Central Tehran
Branch, Islamic Azad University, Tehran, Iran}

\begin{abstract}

From the group theoretical point of view, it is proved that the
theory of linear conformal gravity should be written in terms of a
tensor field of rank-3 and mixed symmetry [Binegar \emph{et al}.,
Phys. Rev. D 27, (1983) 2249]. We obtained such a field equation
in de Sitter space [Takook \emph{et al}, J. Math. Phys. 51, (2010)
032503]. In this paper, a proper solution to this equation is
obtained as a product of a generalized polarization tensor and a
massless scalar field and then the conformally invariant two-point
function is calculated. This two-point function is de Sitter
invariant and free of any pathological large-distance behavior.

\end{abstract}
\maketitle

\section{Introduction}

One of the most important goals of theoretical physics is to
achieve a proper theory of quantum gravity. Attempts of quantizing
General Relativity (GR) have met with some difficulties. Firstly,
the principles of general covariance and causality of GR are in
conflict with quantum states of conventional quantum field theory
since the two principles of GR are closely related to locality but
the quantum states are defined globally. Therefore in order to
quantize GR, the quantum states or the probability amplitudes must
be defined locally in such a case they are compatible with the
principles of GR.

Secondly, the gravitational field is long range and seems to
travel with the speed of light, thus, its quanta (the graviton, if it
exists) should be massless and should propagate on the light cone.
Therefore in the first approximation at least, its equation is
expected to be conformal invariant. Note that equations of
massless particles are conformally invariant (CI) \cite{3},
whereas Einstein's equations are not.

Consideration of massless spin-2 particle is of great importance
since it is among the central objects in quantum cosmology and
quantum gravity. In Minkowski space, the massless field equations
are conformally invariant and for every massless representation of
Poincar\'e group there exists only one corresponding
representation in the conformal group \cite{barut,angelo}.
Therefore, it seems that we need a theory which remains invariant
under conformal transformation (as one expected for massless
theories), and this theory should also be invariant under its
spacetime symmetry group. [According to group representation
theory and its Wigner interpretation in terms of elementary
systems, a linear gravitational field should transform according
to unitary irreducible representations of its spacetime symmetry
group.] Barut and B\"{o}hm \cite{barut} have shown that for the
physical representation of conformal group, the value of conformal
Casimir operator is $9$. But based on work of Binegar et al
\cite{binegar}, for any tensor field of rank-2, this value becomes
8. Therefore, the tensor field of rank-2 does not correspond to
any unitary irreducible representation (UIR) of conformal group.
In other words, the tensor field which corresponds to the physical
representations of the conformal group must be a tensor field of
higher rank.\\ We extended this group theoretical content to the
de Sitter space and used a mixed symmetry tensor field of rank-3
with conformal degree zero, which transforms according to both
UIRs of the conformal and de Sitter groups \cite{takook}. By mixed
symmetry, we mean $$
\Psi_{abc}=-\Psi_{bac},\,\,\,\,\sum_{cycl}\Psi_{abc}=0,$$ while a
field of conformal degree zero satisfies
$u^d\partial_d\Psi_{abc}=0$.

In this paper, we first obtain the solution to this conformal
field equation and then the conformally invariant two-point
function is calculated in such a way that de Sitter invariance is
preserved and the theory is free of pathological large-distance
behavior.

Here, we would like to mention that in the context of linear
quantum gravity, it had been proved that the graviton propagator
in the linear approximation for largely separated points has a
pathological behavior (infrared divergence) and also in de Sitter
(dS) background, the dS invariance is broken \cite{allen,
floratos, antoniadis}. Some authors have suggested that infrared
divergence might lead to instability of the dS space \cite{ford,
antoni1}.Accordingly, Tsamis and Woodard have studied a field
operator for linear gravity in dS space in terms of flat
coordinates \cite{tsamis}. Noting that this coordinate covers only
one-half of the dS hyperboloid, they have examined the possibility
of quantum instability and have found a quantum field, which
breaks dS invariance. However, Antoniadis, Iliopoulos and Tomaras
\cite{antoni} showed that the infrared divergence of the graviton
propagator in one loop approximation is gauge dependent, so it
should not appear in an effective way as a physical quantity; this
later was verified by others \cite{higuchi,vega}. Interestingly,
it is shown that in indefinite metric field quantization (Krein
space method), these two problems are solved. The dS invariance is
survived in Krein space setup as long as a Gupta-Bleuler like
vacuum is used to calculate the physical graviton two-point
function \cite{takook2,dehghani}. And the singularity of the
Wightman two-point function [which appears because of the zero
mode problem of the Laplace-Beltrami operator on dS space
\cite{allenfolacci}] is removed when prescription of the
completely covariant quantization of the minimally coupled scalar
field is followed in Krein space \cite{gazeaurenaud}.

The organization of this paper and its brief outlook are as
follows: section II is devoted to a brief review of the CI
massless spin-2 wave equations in dS space. The solution of the
field equation is considered in section III. It is shown that this
solution can be written in terms of a polarization tensor and a
massless scalar field as
$$ \K_{\alpha\beta}(x)= {\cal D}_{\alpha\beta}(x,\partial)\phi (x). $$
In section IV, the CI bi-tensor two-point function ${\cal
W}_{\alpha\beta \alpha'\beta'}(x,x')$ has been calculated in terms
of a scalar two-point function, $\cal W({\cal{Z}})$. This scalar
two-point function plays a central role in obtaining the conformal
graviton two-point function, we find this two-point function in
indefinite metric field (Krein) quantization method in section V.
Finally a brief conclusion and an outlook for further
investigation has been presented. We have supplied some useful
mathematical details of calculations in the appendices.

\setcounter{equation}{0}
\section{de Sitter field equation}
Astrophysical data coming from type Ia supernova indicate that our
universe is accelerating and can be well approximated by a world
with a non-zero positive cosmological constant \cite{riess}. It
means that our universe, in the first approximation, might be in a
dS phase. de Sitter space plays an essential role in the
inflationary scenario \cite{linde} and also its metric becomes
important at large-scale universe, since the existence of such
non-vanishing positive cosmological constant is proposed to
explain the luminosity observations of the farthest supernovas
\cite{perlmutter}. Thus the quantization of the massless spin-2
field in dS space, without infrared divergence presents an
excellent modality for further research and also it can be an
important element in our understanding of quantum gravity and
quantum cosmology. Let us first review de Sitter space.

\subsection{de Sitter space:}
de Sitter space can be identified by a 4-dimensional hyperboloid
embedded in 5-dimensional flat spacetime:
\begin{equation}
X_H=\{x \in \R^5 ;x^2=\eta_{\alpha\beta} x^\alpha x^\beta
=-H^{-2}=-\frac{3}{\Lambda}\},\;\;\alpha,\beta=0,1,2,3,4,
\end{equation}
where $\eta_{\alpha\beta}=$ diag$(1,-1,-1,-1,-1)$ and $H$,
$\Lambda$ are the Hubble parameter and cosmological constant
respectively. The dS metric is
$$ds^2=\eta_{\alpha\beta}dx^{\alpha}dx^{\beta}=g_{\mu\nu}^{dS}dX^{\mu}dX^{\nu},\;\;\mu,\nu=0,1,2,3$$
where $X^\mu$'s are  $4$ spacetime intrinsic coordinates of the dS
hyperboloid. Any geometrical object in this space can be written
either in terms of four local coordinates $X^\mu$ (intrinsic space
notation) or five global coordinates $x^\alpha$ (ambient space
notation).

Kinematical group of the dS space is the $10$-parameter group
$SO_0(1,4)$ (connected component of the identity in $O(1,4)$),
which is one of the two possible deformations of the Poincar\'e
group. There are two Casimir operators, \begin{equation}
Q^{(1)}_2=-\frac{1}{2}L^{\alpha\beta}L_{\alpha\beta},\;\;\;\;\
Q^{(2)}_2=-W_{\alpha}W^{\alpha},\end{equation} where
$W_{\alpha}=-\frac{1}{8}\epsilon_{\alpha\beta\gamma\sigma\eta}L^{\beta\gamma}L^{\sigma\eta},$
with 10 infinitesimal generators $
L_{\alpha\beta}=M_{\alpha\beta}+S_{\alpha\beta}.$ The subscript
$2$ in $Q^{(1)}_2$, $Q^{(2)}_2$ reminds us that the carrier space
is constituted by second rank tensors. $M_{\alpha\beta}$ and
$S_{\alpha\beta}$ are the orbital and the spinorial parts
respectively \cite{gazeauhans}. The symbol
$\epsilon_{\alpha\beta\gamma\sigma\eta}$ holds for the usual
antisymmetric tensor.  "Massless"\footnote{Note that in dS space,
concept of mass does not exist by itself as a conserved quantity.
The term "massive" is referred to fields that in their zero
curvature limit reduce to massive Minkowskian fields \cite{barut}.
The Concept of light-cone propagation, however, does exist and
leads to the conformal invariance.} is used in reference to
propagation on the dS light cone (conformal invariance). The
conformal invariance and the light-cone propagation, constitute
the basis for constructing massless field in dS space. As a matter
of fact, we address the massless spin-2 field in dS space, to one
kind of representation, namely the lowest representation of rank-2
tensor in discrete series of the dS group. According to the de
Sitter group, massless spin-2 field is denoted by
$\Pi^{\pm}_{2,2}$ and $\Pi^{\pm}_{2,1}$ in which $\Pi_{p,q}$'s are
UIRs of the dS group in its discrete series and the sign $\pm,$
stands for the helicity. The pair $(p,q)$ is used to label the
UIRs in de Sitter group. It is proved that $\Pi^{\pm}_{2,2}$, have
a Minkowskian interpretation.

The compact subgroup of conformal group $SO(2,4)$ is $SO(2)\otimes
SO(4)$. Let $C(E;j_1,j_2)$ denote the irreducible projective
representation of the conformal group, where $E$ is the
eigenvalues of the conformal energy generator of $SO(2)$ and
$(j_1,j_2)$ is the $(2j_1+1)(2j_2+1)$ dimensional representation
of $SO(4)=SU(2)\otimes SU(2)$. The representation $\Pi^+_{2,2}$
has a unique extension to a direct sum of two UIRs $C(3;2,0)$ and
$C(-3;2,0)$ of the conformal group, with positive and negative
energies respectively \cite{barut,levy}. The latter restricts to
the massless Poincar\'e UIRs $P^>(0, 2)$ and $P^<(0,2)$ with
positive and negative energies respectively. $ {\cal P}^{
\stackrel{>} {<}}(0,2)$ (resp. $ {\cal
P}^{\stackrel{>}{<}}(0,-2)$)  are the massless Poincar\'e UIRs
with positive and negative energies and positive (resp. negative)
helicity. The following diagrams illustrate these connections
\begin{equation}
\left.
\begin{array}{ccccccc}
&& {\cal C}(3,2,0)& &{\cal C}(3,2,0)&\hookleftarrow &{\cal P}^{>}(0,2)\\
\Pi^+_{2,2} &\hookrightarrow &\oplus&\stackrel{H=0}{\longrightarrow} & \oplus  & &\oplus\\
&& {\cal C}(-3,2,0)& & {\cal C}(-3,2,0) &\hookleftarrow &{\cal
P}^{<}(0,2),\\
\end{array}
\right.
\end{equation}

\begin{equation}
\left.
\begin{array}{ccccccc}
&& {\cal C}(3,0,2)& &{\cal C}(3,0,2)&\hookleftarrow &{\cal P}^{>}(0,-2)\\
\Pi^-_{2,2} &\hookrightarrow &\oplus&\stackrel{H=0}{\longrightarrow}&\oplus &&\oplus\\
&& {\cal C}(-3,0,2)&& {\cal C}(-3,0,2)&\hookleftarrow &{\cal P}^{<}(0,-2),\\
\end{array}
\right.
\end{equation} where the arrows $\hookrightarrow $ designate unique
extension. It is important to note that the representations
$\Pi^{\pm}_{2,1}$ do not have corresponding flat limit.
Mathematical details of the group contraction and the physical
principles underlying the relationship between dS and Poincar\'e
groups can be found in Refs. \cite{levy} and \cite{bacry},
respectively.

\subsection{Dirac's six cone formalism and conformal-invariant field equations:}

The conformal group acts nonlinearly on Minkowski coordinates.
Dirac proposed a manifestly conformally covariant formulation in
which the Minkowski coordinates are replaced by coordinates on
which the conformal group acts linearly. The resultant theory is
then formulated on a 5-dimensional hypercone (named Dirac's
six-cone) in a 6-dimensional space. This method was first used by
Dirac \cite{dirac} to demonstrate the field equations for spinor
and vector fields in $(1+3)$-dimensional spacetime in a manifestly
CI form. This approach to conformal symmetry which leads to best
path to exploit the physical symmetry was then developed by Mack
and Salam \cite{mack} and many others \cite{kastrup}.

Dirac's six-cone, or Dirac's projection cone, is defined by \b
u^2\equiv (u^0)^{2}-\vec u^{2}+(u^5)^{2}=\eta_{ab} u^a u^b=0 ,\;\;
\eta_{ab}=\mbox{diag}(1,-1,-1,-1,-1,1),\e where $ \;u^{a} \in
\R^{6},$ and  $ \vec u \equiv(u^{1},u^{2},u^{3},u^{4})$. Reduction
to four dimensions is achieved by projection, that is by fixing
the degrees of homogeneity of all fields. Wave equations,
subsidiary conditions, etc., must be expressed in terms of
operators that are defined intrinsically on the cone. These are
well-defined operators that map tensor fields to tensor fields
with the same rank on the cone $u^2=0$. So, the resultant
equations which are obtained by this method, are conformally
invariant.

We studied this method in de Sitter space and obtained the field
equations for massless scalar and vector fields
\cite{takook,dehghani,behroozi}. It has been shown that in the
flat limit $(H \rightarrow 0)$, these CI equations, reduce exactly
to their counterpart in Minkowsi space, e.g., Maxwell equations
are obtained from the vector field case \cite{dehghani,behroozi}.
The mixed symmetric tensor field $F_{\alpha\beta\gamma}$ in dS
space is defined for the spin-2 case which is related to the
rank-2 field ${\cal K}_{\alpha\beta}$ via the following relation:
\b \label{2} F_{\alpha\beta\gamma} \equiv\Big(\bar
\partial_\alpha +x_\alpha\Big)\K_{\beta\gamma}-\Big(\bar
\partial_\beta +x_\beta
\Big)\K_{\alpha\gamma},\e where $\bar\partial$ is a transverse
derivative (Appendix A). For the spin-2 case it is found that
\cite{takook}, \b \label{equ}
(Q_0-2)^2Q_0\K_{\beta\gamma}=0,\,\,\,\mbox{or equivalently},\,\,
(Q_{2}+4)^2(Q_{2}+6){\cal K}_{\alpha\beta}=0,\e where $Q_2\equiv
Q^{(1)}_2 $, and
$Q_0=-{\frac{1}{2}M_{\alpha\beta}M^{\alpha\beta}}$ (Appendix A).

We like to emphasize that since Dirac's six-cone formalism has
been used, Eq. (\ref{equ}) is CI, and also leads to the UIR of the
dS and conformal groups. At the next stage we will obtain the
solution of Eq. (\ref{equ}).

\setcounter{equation}{0}
\section{de Sitter field solution}

Let us start with the most generic form of ${\cal
\K}_{\alpha\beta}$ which can be chosen as \cite{garidigatak} \b
{\cal \K}_{\alpha\beta}=\theta_{\alpha\beta}\phi_1+ {\cal S}\bar
Z_{1\alpha}K_\beta+D_{2\alpha}K_{g\beta},\e where ${\cal S}$ is
the symmetrizer operator and $ Z_{1} $ is a constant 5-dimensional
vector, $ \phi_1 $ is a scalar field, $ K $ and $ K_g $ are two
vector fields. Bar over the vector makes it a tangential (or
transverse) vector on dS space,$$ \bar
Z_{\alpha}=\theta_{\alpha\beta}Z^{\beta}=Z_{\alpha}+H^2x_{\alpha}x\cdot
Z,\,\,\, \mbox{with}\,\,\,x\cdot\bar Z\equiv x_\alpha
\bar{Z}^\alpha=0\,,$$ and $\theta_{\alpha\beta}$ is the transverse
projector
($\theta_{\alpha\beta}=\eta_{\alpha\beta}+H^2x_{\alpha}x_{\beta}\,$).
The operator $D_2$ is the generalized gradient defined by (for
simplicity from now on we take $H$=1)
$$D_2K={\cal S}(\bar\partial-x)K.$$
If ${\cal \K}_{\alpha\beta}$ satisfies the divergenceless and
transversality conditions (which are needed in order to relate it
to the physical representation) then, after doing some easy
algebra, one gets:
\begin{equation}
\begin{aligned} \label{phi1}
{\cal K}'=0,\hspace{3mm}x\cdot (
K\,\,\mbox{and}\,\,K_g)=&\bar\partial
\cdot K=0,\\
2\phi_1+Z_1.K+\bar{\partial}.K_g=0.
\end{aligned}
\end{equation}
Substituting $ \K_{\alpha\beta} $ in (\ref{equ}) results in

\begin{equation}\label{ar}
\left\{\begin{array}{rl} (Q_0+4)^2(Q_0+6)\phi_1=&-4[
Q_1(Q_1+2)+(Q_0+4)(Q_1+2)+(Q_0+4)^2 ]Z_1.K,\,\,\,\,\,\,
{(I)}\vspace{2mm}\\\vspace{2mm}
Q_1^2(Q_1+2)K =&0, \hspace{8.7cm} {(II)}\\
(Q_1+4)^2(Q_1+6)K_g=&2[(x.Z_1)Q_1(Q_1+2)+(Q_1+4)(x.Z_1)(Q_1+2)\\
&\,\,\,\,\,\,+(Q_1+4)^2(x.Z_1)] K.
\hspace{5cm}{(III)}\end{array}\right.
\end{equation}
It is easy to show that from relations (\ref{ar}-$I$,
\ref{ar}-$II)$ together with the conditions given in (\ref{phi1}),
one obtains

\b \label{phi12} \phi_1=-\frac{2}{3}Z_1.K,
\;\;\;\;\;\;Q_0(Q_0-2)^2 \phi_1=0, \e note that the latter is the
massless scalar field equation in dS space
\cite{gazeaurenaud,behroozi}. On the other hand, from
(\ref{phi12}) and (\ref{phi1}), we find \b \bar{\partial}\cdot K_g
=\frac{1}{3}Z_1.K.\e What has been done up to now is to write
$\phi_1$ and $K_g$ in terms of $K,$ now we want to obtain $K.$

 $K$ is a vector field that satisfies the conditions (\ref{phi1}), it can be written as \cite{garidigarou, gazeautakook}
\b\label{K} K=\bar Z_2 \phi_2+D_1 \phi_3,\e where $ Z_2 $ is
another 5-dimensional constant vector, $ \phi _2 $ and $ \phi_3 $
are two scalar fields and $D_1=\bar\partial$. Substituting $K$
into (\ref{ar}-$II)$ results in \b \label{3.8} Q_0(Q_0-2)^2 \phi_2
=0, \e it is interesting to note that $\phi_2$ also satisfies
massless field equation. Similarly, $\phi_3$ can be written in
terms of $\phi_2$ as follows (Appendix B) \b \label{3.9}\phi_3
=-[(x.Z_2)+(Z_2.\bar\partial)]\phi_2,\e and from (\ref{K},
{\ref{phi12}), one obtains
\begin{equation}
\begin{aligned} \label{3.10}
K=&\Big(\bar
Z_{2}-D_1[(x.Z_2)+(Z_2.\bar\partial)]\Big)\phi_2,\\
\phi_1 =&-\frac{2}{3}{Z_1}.\bigg[\bar
Z_{2}-D_1[(x.Z_2)+(Z_2.\bar\partial)]\bigg]\phi_2.
\end{aligned}
\end{equation}

From Eq. (\ref{ar}-$III)$ and after making use the similar procedure
given in Appendix B, it is proved that $K_g$ can be written in
terms of $K$ as
\begin{equation} \label{3.11}
K_g=\frac{1}{3}\left[4(x.Z_1)K+Z_1.\bar\partial K -x(Z_1.K)
\right],
\end{equation}
note that $ x.K_g=0$ and $\bar{\partial}.K_g =\frac{1}{3}Z_1.K$.

Gathering all the results and from Eqs. (\ref{3.10}) and
(\ref{3.11}), we can construct the tensor field
$\K_{\alpha\beta}$, in terms of a massless scalar field as
follows: \b \label{equat} \K_{\alpha \beta}(x)={\cal D}_{\alpha
\beta}(x,\partial,Z_1,Z_2)\phi_2,\e where ${\cal D}$ is the
projector tensor defined by
\begin{equation}
\begin{aligned} {\cal
D}(x,\partial,Z_1,Z_2)=&\\
\bigg[-\frac{2}{3}& \theta Z_1.+{\cal S}\bar Z_1+\frac{1}{3}D_2
\left[4(x.Z_1)+Z_1.\bar\partial -x(Z_1.) \right]\bigg]\bigg[ \bar
Z_{2}-D_1[(x.Z_2)+(Z_2.\bar\partial)]\bigg].
\end{aligned}
\end{equation}

It is more suitable to express tensor field (\ref{equat}) in terms
of a polarization tensor and de Sitter plane wave and then by
taking the flat limit one can fix $Z_1$ and $Z_2$. This can be
achieved by written $\phi_2$ as a de Sitter plane wave \cite{bros,
bros2} \b \phi_2(x)=(Hx.\xi)^l ,\e where $\xi \in \R^5 $ lies on
the positive null cone $ {\cal C}^{+} = \{ \xi \in \R^5;\;\;
\xi^2=0,\; {\xi}^0>0 \}$. Therefore equation (\ref{equat}) can be
easily brought into the following form:
$$ \K_{\alpha \beta}(x)= {\cal
E}_{\alpha\beta}(x,\xi,Z_1,Z_2)(Hx.\xi)^l,$$ where ${\cal E}$ is
the generalized polarization tensor. Now one can fix the two
arbitrary constant vector $Z_1$ and $Z_2$ in terms of the
polarization tensor of massless spin-2 field in the Minkowskian
limit \cite{garidigarou}. In which, $l=-1,-2$ and $l=0,-3$ leads
to conformally coupled and minimally coupled massless scalar
fields in de Sitter space, respectively \cite{bros, bros2}.

\setcounter{equation}{0}
\section{Two-point function}

The two-point functions in de Sitter space can be written in terms
of bi-tensors \cite{allen2}. These are functions of two points
$(x, x')$ which behave like tensors under coordinate
transformations at each points. Bi-tensors are called maximally
symmetric if they respect de Sitter invariance. Furthermore, as
explained in \cite{bros2} and \cite{garidigarou}, the axiomatic
field theory in de Sitter is based on bi-tensor two-point
function. This two-point function is defined by \b {\cal
W}_{\alpha\beta \alpha'\beta'}(x,x')=\langle
\Omega|\K_{\alpha\beta}(x)\K_{\alpha'\beta'}(x')|\Omega  \rangle
,\e where $x,x'\in X_H$ and $|\Omega\rangle $ is the Fock-vacuum
state. The two-point function which is a solution of Eq.
(\ref{equ}) with respect to $x$ and $x'$, can be written in terms
of a scalar two-point function as
$$ {\cal
W}_{\alpha\beta\alpha'\beta'}(x,x')=\Delta_{\alpha\beta\alpha'\beta'}{\cal{W}}(x,x'),$$
where $ {\cal W}(x,x')$ and $\Delta_{\alpha\beta\alpha'\beta'}$,
are bi-scalar two-point function and bi-tensor projection
operator, respectively.

\subsection{Two-point function in ambient space notation:}

The similar procedure as the previous section is used to obtain
the transverse two-point function, therefore one can write
\b\label{4.2} {\cal W}_{\alpha\beta
\alpha'\beta'}(x,x')=\theta_{\alpha\beta}
\theta'_{\alpha'\beta'}{\cal W}_0(x,x')+{\cal S}{\cal
S}'\theta_{\alpha}.\theta'_{\alpha'}{
W}_{1\beta\beta'}(x,x')+D_{2\alpha}D'_{2\alpha'}{
W}_{g\beta\beta'}(x,x'),\e note that $D_2D'_2=D'_2D_2$ and ${
W}_{1}$ and ${ W}_{g}$ are transverse bi-vector two-point
functions which will be identified later. At this stage it is
shown that the calculation of ${\cal W}_{\alpha\beta
\alpha'\beta'}(x,x')$ could be initiated from $x$ or $x'$, without
any difference, this means each choice results to the same
equation for ${\cal W}_{\alpha\beta \alpha'\beta'}(x,x')$. With
the choice of $x$, ${\cal W}_{\alpha\beta \alpha'\beta'}(x,x')$
must satisfy Eq. $(\ref{equ})$, it is a matter of simple
calculation to get the following relations:
\begin{equation}\label{ar2}
\left\{\begin{array}{rl} (Q_0+4)^2(Q_0+6)\theta'{\cal
W}_0&=-4{\cal S}'[ Q_1(Q_1+2)+(Q_0+4)(Q_1+2)+(Q_0+4)^2 ]
\theta'.{W}_{1},\;\;\;\;\;\;{(I)}\vspace{2mm}\\
Q_1^2(Q_1+2){ W}_{1}&=0, \hspace{8cm}{(II)}\vspace{2mm}\\
(Q_1+4)^2(Q_1+6)D'_2{ W}_g&=2{\cal
S}'[(x.\theta')Q_1(Q_1+2)+(Q_1+4)(x.\theta')(Q_1+2)\\
&\,\,\,\,\,\,\,\,\,\,\,\,\,\,\,\,\,\, +(Q_1+4)^2(x.\theta')] {
W}_{1}.\hspace{4cm}{(III)}
\end{array}\right.
\end{equation}
Noting that $W_1$ is divergenceless, Eq. (\ref{ar2}-$I)$ implies
that \b \label{4.4} \theta'{\cal W}_0(x,x')=-\frac{2}{3}{\cal
S}'\theta'.{ W}_{1}(x,x').\e In order to handle Eq.
(\ref{ar2}-$II)$, we write ${W}_1$ in terms of two bi-scalar
two-point functions as follows
$${W}_{1}=\theta.\theta'{\cal W}_{2}+D_1D'_1{\cal W}_{3}.$$
Substituting ${W}_{1}$ in Eq. (\ref{ar2}-$II)$ and using the
divergenceless condition, one obtains
\begin{equation}
\begin{aligned}
&D'_1{\cal W}_{3}=-[x.\theta'{\cal W}_2+\theta'.\bar\partial{\cal
W}_{2}],\\
&Q_0(Q_0-2)^2{\cal W}_{2}=0.
\end{aligned}
\end{equation}
In the next section we will consider ${\cal W}_{2}$ in more
detail. Setting ${\cal W}_{2}\equiv \cal W,$ one obtains
\b\label{4.6} { W}_{1}(x,x')=\left(\theta.\theta'
    -D_{1}[x.\theta'+\theta'.\bar\partial]\right){\cal W}(x,x'),\e then Eq. $(\ref{ar2}-III)$
leads to
 \b\label{4.7} D'_2{ W}_g(x,x')=\frac{1}{3}{\cal{S}'}\left[ 4(x.{\theta}') + ({\theta'}.{\bar\partial}) - x({\theta'}.)\right]{W}_{1}(x,x').\e Now,
 we use Eqs. $(\ref{4.4})$, $(\ref{4.6})$ and
$(\ref{4.7})$, to write ${\cal W}_{\alpha\beta
\alpha'\beta'}(x,x')$ as \b {\cal W}_{\alpha\beta
\alpha'\beta'}(x,x')=\Delta_{\alpha\beta \alpha'\beta'}
(x,\partial,x',\partial'){\cal W}(x,x'),\label{twopoint} \e where

\begin{equation}
\begin{aligned}
\Delta_{\alpha\beta
\alpha'\beta'}(x,\partial,x',\partial')=&-\frac{2}{3}{\cal
S'}\theta \theta'.\Big(\theta.\theta'
    -D_{1}[x.\theta'+\theta'.\bar\partial]\Big)
 +{\cal S}{\cal S}'\theta.\theta'\Big(\theta.\theta'
    -D_{1}[x.\theta'+\theta'.\bar\partial]\Big)\\
     +\frac{1}{3}& D_2{\cal
S}'\Big( 4(x.{\theta}') + ({\theta'}.{\bar\partial}) -
x({\theta'}.)\Big)\Big(\theta.\theta'
    -D_{1}[x.\theta'+\theta'.\bar\partial]\Big)
\end{aligned}
\end{equation}
Similarly, with the choice of $x'$, the two-point function
(\ref{4.2}) satisfies Eq. (\ref{equ}) (with respect to $x'$, see
Appendix A).

In some cases of interest it is useful to express the relations in
terms of $ {\cal{Z}}$ which is an invariant object under the
isometry group $O(1,4)$. It is defined for two given points on the
dS hyperboloid $x$ and $x'$, by $$ {\cal{Z}}\equiv
-x.x'=1+{\frac{1}{2}}(x-x')^2,$$ note that any function of ${\cal
Z}$ is dS invariant, as well. It is the work of a few lines to
show that (\ref{twopoint}) in terms of $ {\cal{Z}}$ becomes:

\begin{equation}
\begin{aligned} \label{4.13}
{\cal W}_{\alpha\beta \alpha'\beta'}(x,x')&= {\frac{1}{3}}{\cal
S}{\cal
S}'\bigg[\theta_{\alpha\beta}\theta'_{\alpha'\beta'}f_1({\cal{Z}})+(\theta_{\alpha}.\theta'_{\alpha'})(\theta_{\beta}.\theta'_{\beta'})f_2({\cal{Z}})
+\theta'_{\alpha'\beta'}(x'.\theta_{\alpha})
(x'.\theta_{\beta})f_3({\cal{Z}})+\\
(x'.\theta_{\alpha})(x'.\theta_{\beta})(x.\theta'_{\alpha'})&(x.\theta'_{\beta'})f_4({\cal{Z}})
+(\theta_{\alpha}.\theta'_{\alpha'})(x.\theta'_{\beta'})(x'.\theta_{\beta})f_5({\cal{Z}})
+\theta_{\alpha\beta}(x.\theta'_{\alpha'})(x.\theta'_{\beta'})
f_6({\cal{Z}})\bigg]\frac{d}{d{\cal{Z}}}{\cal W}({\cal{Z}}),
\end{aligned}
\end{equation}
in which

$$f_1({\cal Z})=-{\cal Z}(1+{{\cal{Z}}}{\frac{d}{d{\cal{Z}}}}),\,\,
f_2({\cal{Z}})=-{\cal{Z}}(11+2{{\cal{Z}}}{\frac{d}{d{\cal{Z}}}}),\,\,
f_3({\cal{Z}})=-(3+{{\cal{Z}}}{\frac{d}{d{\cal{Z}}}})\frac{d}{d{\cal{Z}}},$$
$$f_4({\cal{Z}})=-(18+10{\cal{Z}}{\frac{d}{d{\cal{Z}}}}+{{\cal{Z}}^2}{\frac{d^2}{d{\cal{Z}}^2}})\frac{d}{d{\cal{Z}}},\,\,f_5({\cal{Z}})=
(34+31{\cal
Z}\frac{d}{d{\cal{Z}}}+4{{\cal{Z}}^2}{\frac{d^2}{d{\cal{Z}}^2}}),$$
$$
\,\,f_6({\cal{Z}})=-{\cal{Z}}(15+9{\cal{Z}}{\frac{d}{d{\cal{Z}}}}+{{\cal{Z}}^2}{\frac{d^2}{d{\cal{Z}}^2}}).$$
This form of two-point function satisfies the traceless and
divergenceless conditions:
$$\bar{\partial}.{\cal
W}=\bar{\partial'}.{\cal W}=0,\;\;\;\mbox{and}\;\;\;{\cal
W}_{\alpha\beta \alpha'}^{\;\;\;\;\alpha'}(x,x')={\cal
W}^{\alpha}_{\alpha \alpha'\beta'}(x,x')=0.$$

\subsection{Two-point function in intrinsic space notation:}

The two-point function (\ref{4.13}), has been written in ambient
space and here we want to project this two-point function to the
intrinsic space. It is shown that any maximally symmetric
bi-tensor can be expressed as a sum of products of three basic
tensors \cite{allen2} whose coefficients are functions of the
geodesic distance and parallel propagator which is defined by
$$n_\mu = \nabla_\mu \sigma(x, x')\;\;\;,\;\;\; n_{\mu'} = \nabla_{\mu'} \sigma(x,
x'),$$
$$g_{\mu\nu'}=-c^{-1}({\cal{Z}})\nabla_{\mu}n_{\nu'}+n_\mu n_{\nu'}.$$
The geodesic distance is implicitly defined for $
{\cal{Z}}=-x\cdot x', $ by \b
\left\{%
\begin{array}{ll}
     {\cal{Z}}=\cosh (\sigma ), & \hbox{if $x$ and $x'$ are
time-like separated;} \\
     {\cal{Z}}=\cos (\sigma ), & \hbox{if
$x$ and $x'$are space-like separated.} \\
\end{array}%
\right.\e The basic bi-tensors in ambient space notation are
obtained
$$ \bar{\partial}_\alpha \sigma(x,x')\;\;\;,\;\;\;\bar{\partial}'_{\beta'}
\sigma(x,x')\;\;\;,\;\;\;\theta_\alpha .\theta'_{\beta'},$$ which
are restricted to the hyperboloid by
$$ {\cal{T}}_{\mu\nu'}=\frac{\partial x^\alpha}{\partial
X^\mu}\frac{\partial x'^{\beta'}}{\partial
X'^{\nu'}}T_{\alpha\beta'}.$$

For $ {\cal{Z}}=\cos(\sigma), $ one can find
$$n_\mu=\frac{\partial x^\alpha}{\partial X^\mu}\bar{\partial}_\alpha \sigma(x,x')=
\frac{\partial x^\alpha}{\partial X^\mu} \frac{(x' \cdot
\theta_\alpha)}{\sqrt{1-{\cal{Z}}^2}},\;\; n_{\nu'}=\frac{\partial
x'^{\beta'}}{\partial X'^{\nu'}}\bar{\partial}_{\beta'}
\sigma(x,x') =\frac{\partial x'^{\beta'}}{\partial X'^{\nu'}}
\frac{(x\cdot\theta'_{\beta'})}{\sqrt{1-{\cal{Z}}^2}},$$
$$\nabla_\mu n_{\nu'}=\frac{\partial x^\alpha}{\partial
X^\mu}\frac{\partial x'^{\beta'}}{\partial
X'^{\nu'}}\theta^\varrho_\alpha
\theta'^{\gamma'}_{\beta'}\bar{\partial}_\varrho\bar{\partial}_{\gamma'}
\sigma(x, x')=c({\cal{Z}})\Big[n_\mu
n_{\nu'}{\cal{Z}}-\frac{\partial x^\alpha}{\partial
X^\mu}\frac{\partial x'^{\beta'}}{\partial X^{\nu'}}\theta_\alpha
\cdot\theta'_{\beta'}\Big],$$ where $
c^{-1}({\cal{Z}})\equiv-\frac{1}{\sqrt{1-{\cal{Z}}^2}}.$  \\For $
{\cal{Z}}=\cosh (\sigma), $  $ n_\mu $ and $ n_\nu$ are multiplied
by $i$ and then $ c({\cal{Z}}) $ becomes
$-\frac{i}{\sqrt{1-{\cal{Z}}^2}}.$ In both cases we have
$$ g_{\mu\nu'}+({\cal{Z}}-1)n_\mu n_{\nu'}=\frac{\partial x^\alpha}{\partial
X^\mu}\frac{\partial x'^{\beta'}}{\partial X'^{\nu'}}\theta_\alpha
\cdot\theta'_{\beta'} .$$

Similarly, the two-point functions in ambient space are related to
those in de Sitter intrinsic space through
$$ Q_{\mu\nu\mu'\nu'}= \frac{\partial x^\alpha}{\partial X^\mu}
\frac{\partial x^\beta}{\partial X^\nu} \frac{\partial
x'^{\alpha'}}{\partial X'^{\mu'}} \frac{\partial
x'^{\beta'}}{\partial
X'^{\nu'}}{\cal{W}}_{\alpha\beta\alpha'\beta'}.$$ Finally, the
resultant two-point function in dS intrinsic space reads
$$
Q_{\mu\nu\mu'\nu'}(X,X')={\frac{(1-{\cal{Z}}^2)^2}{3}}{\cal
S}{\cal
S}'\left[\;g_{\mu\nu}g'_{\mu'\nu'}\;\frac{f_1}{(1-{\cal{Z}}^2)^2}
+g_{\mu\mu'}g_{\nu\nu'}\;\frac{f_2}{(1-{\cal{Z}}^2)^2}\right.$$
$$+\;g'_{\mu'\nu'}n_{\mu}
n_{\nu}\;\frac{f_3}{1-{\cal{Z}}^2}+g_{\mu\mu'}n_\nu
n_{\nu'}\left(\frac{2({\cal{Z}}-1)f_2}{(1-{\cal{Z}}^2)^2}
+\frac{f_5}{1-{\cal{Z}}^2} \right)
$$
\b\label{two2}\left.+n_{\mu}
n_{\nu}n_{\mu'}n_{\nu'}\left(\frac{f_2}{(1+{\cal{Z}})^2}-\frac{f_5}{1+{\cal{Z}}}+f_4
\right)+g_{\mu\nu}n_{\mu'}n_{\nu'}\;\frac{f_6}{1-{\cal{Z}}^2}
\right]{\cal W}({\cal{Z}}),\e  in the next section, ${\cal
W}({\cal{Z}})$  is calculated.

\setcounter{equation}{0}
\section{Scalar field two-point function}

In the previous section, ${\cal W}_{\alpha\beta\alpha'\beta'}$ was
calculated in terms of a scalar field two-point function, ${\cal
W}({\cal{Z}})$, that satisfies $ Q_0(Q_0-2)^2{\cal
W}({\cal{Z}})=0.$ Now, we want to obtain this scalar two-point
function. Let us write the general statement of ${\cal
W}({\cal{Z}})$ as follows:
\begin{equation} {\cal W}({\cal{Z}})=
c_1A({\cal{Z}})+c_2B({\cal{Z}})+c_3C({\cal{Z}})+c_4D({\cal{Z}})+c_5E({\cal{Z}}),\end{equation}
where $c_1$, $c_2$, $c_3$, $c_4$, $c_5$ are constants and each one
of $A({\cal{Z}})$, $B({\cal{Z}})$, $C({\cal{Z}})$, $D({\cal{Z}})$,
$E({\cal{Z}})$ is a part of the answer that satisfies the
following equations:
$$ Q_0(Q_0-2)^2A({\cal{Z}})=0,\;\
Q_0(Q_0-2)B({\cal{Z}})=0,\;\ (Q_0-2)^2C({\cal{Z}})=0,$$
$$(Q_0-2)D({\cal{Z}})=0,\;\ Q_0E({\cal{Z}})=0.$$ Each of these functions should be identified. $E({\cal{Z}})$ can be considered as the two-point function for a
minimally coupled massless scalar field in dS space
\cite{allenfolacci,folacci}. This two-point function has been
found in \cite{folacci} as follows \b
E({\cal{Z}})=\frac{1}{8\pi^2}\left[\frac{1}{1-{\cal{Z}}}-\ln(1-{\cal{Z}})+\ln
2+f(\eta,\eta')\right],\e where $f$ is a function of the conformal
time $\eta$ that breaks the dS invariance and because of the term
$\mbox{ln}(1-{\cal Z})$, at largely separated points infrared
divergence appears. However, in Krein space calculation, one
obtains \cite{gazeaurenaud,takook3} \b\label{twopointkrein}
E_K({\cal{Z}})=\frac{i}{8\pi^2}\epsilon
(x^0-x'^{0})\left[\delta(1-{\cal{Z}})+{\vartheta({\cal{Z}}-1)}\right],\e
where ${\vartheta}$ is the Heaviside step function and
\b \epsilon (x^0-x'^0)=\left\{ \ba{rcl} 1&x^0>x'^0 ,\\
0&x^0=x'^0 ,\\ -1&x^0<x'^0.\\ \ea\right.\e Notice that this
two-point function has been written in terms of ${\cal Z}$,
therefore dS invariance is indeed preserved and it is clearly free
of infrared divergence.  $D({\cal{Z}})$ is the two-point function
for a conformally coupled massless scalar field in dS space
\cite{chernikov}: \b D({\cal{Z}})= -{\frac{1}{8\pi^2}}[
{\frac{1}{1-\cal{Z}}}-i\pi\epsilon (x^0-x'^0)\delta(1-{\cal Z})],
\e and in the Krein space, we obtain \cite{gazeaurenaud} \b
D_K({\cal{Z}})= {\frac{i}{8\pi}}\epsilon (x^0-x'^0)\delta(1-{\cal
Z}). \e It is worth noting that $D({\cal{Z}})$ preserves the dS
invariant in both methods of quantization, either in usual way in
Hilbert space or in Krein space quantization method, however,
$E({\cal{Z}})$ is dS invariant only when it is calculated in
indefinite metric field quantization method. In other words, in
order to have a covariant quantization we should carry out the
calculations in Krein space or the Gupta-Bleuler vacuum is needed
for quantization \cite{gazeaurenaud,takook3}. Therefore, we do the
calculations in the Krein space (for the sake of simplicity, we omit the index $K$ for a while and write two-point functions in Krein space).\\
Other functions can be obtained easily by the integration of
$E({\cal{Z}})$ and $D({\cal{Z}})$:
\begin{equation}
\begin{aligned}\label{5.8}
A({\cal{Z}})=&\frac{i}{8\pi^2(1-{\cal{Z}}^2)}\epsilon
(x^0-x'^0)\left[-\frac{1}{4}({\cal{Z}}-1)^2{\vartheta({\cal{Z}}-1)}\right.\\
&\left.+\Big({\vartheta({\cal{Z}}-1)}+{\vartheta(1-{\cal{Z}})}\Big)\Big((\ln|{\cal{Z}}+1|-1)({\cal{Z}}+1)-({\cal{Z}}-1)\ln2-\ln4+2\Big)\right],\\
B({\cal{Z}})=&\frac{i}{8\pi^2(1-{\cal{Z}}^2)}\epsilon
(x^0-x'^0)\left[\frac{1}{2}({\cal{Z}}-1)^2{\vartheta({\cal{Z}}-1)}
+(1-{\cal{Z}}){\vartheta(1-{\cal{Z}})}\right],\\
C({\cal{Z}})=&\frac{i}{8\pi(1-{\cal Z}^2)}\epsilon
(x^0-x'^0)(1-{\cal{Z}}){\vartheta(1-{\cal{Z}})}.
\end{aligned}
\end{equation}
As a result, ${\cal W}({\cal{Z}})$ is obtained in terms of
massless minimally and conformally coupled scalar two-point
functions.

\section{Conclusion}

In the framework of quantum field theory, the graviton is supposed
to be a mediator of the gravitational field. So, if the graviton
exists, it must be a massless spin-2 particle (because the
gravitational field has unlimited range and the source of
gravitation is a second-rank tensor). On the other hand as proved
in \cite{3}, the relativistic equations for massless particles are
invariant under the conformal transformations. \\It was pointed
out that Einstein's theory of gravitation, in the linear
approximation and in the background field method,
$g_{\mu\nu}=g_{\mu\nu}^{BG}+h_{\mu\nu}$, can be considered as a
theory of massless symmetric tensor field of rank-2, however,
contrary to the Maxwell equations (which in the quantum framework,
are regarded as equations of massless spin-1 particle that are
conformally invariant), Einstein's equation of gravitation, as
well as the equation of $h_{\mu\nu}$, is not conformally
invariant. [Notice that conformally invariant equation of the Weyl
gravity in its linear form does not transform according to UIRs of
the background spacetime symmetry group (for example dS group)
\cite{JHEP}.] Moreover, there is no successful theory of quantum
gravity, since the standard theory of gravity is not
renormalizable when quantum gravitational fluctuations are
considered. It has been often claimed that the theories whose
field equations contain higher order derivatives are better to
renormalize than the standard gravity, however, in such theories
one should take care about the unitarity \cite{stelle}. In a
series of papers Bender et.al., have argued that if such theories
are Parity-Time reversal (${\cal PT}$)-symmetric then the
unitarity would survive and they have discussed some higher order
theories which are both renormalizable and unitary \cite{bender}.
In our case the higher order field equation (\ref{equ}) has in
fact ${\cal PT}$-invariance since the Casimir operators of the de
Sitter group are ${\cal PT}$-symmetric. On the other hand it is
worth noting that equation (\ref{equ}) transforms according to the
UIRs of de Sitter group. In this work, we solved this field
equation and found the proper solution in terms of a generalized
polarization tensor and de Sitter plane wave. Then the related CI
two-point functions (Eq. (\ref{4.13}) in ambient space and Eq.
(\ref{two2}) in dS intrinsic space) were obtained. We would like
to emphasize that these two-point functions are invariant under
conformal group as well as de Sitter group.

As is well known, in the theory of quantum fields, Green's
functions are used to study physical quantities. However, in
calculating Green's functions some infinities appear of which most
are removed by the means of regularization and renormalization
procedures. This theory successfully unified the electromagnetic,
weak and strong interactions within the famous so-called Grand
Unified Theory. But up to present days, there is no satisfactory
quantum description of gravity. One needs such a theory to better
understand the influence of the gravitational field on quantum
phenomena or to explain some cosmological observations such as the
anisotropy of the cosmic microwave background radiation
\cite{Rodrigues}. Therefore, over the past fifty years, one of the
great challenges of physics has been the achievement of a proper
theory of the quantized gravitational field.

In the previous work, we had found the field equation for the
massless spin-2 field in de Sitter space which was conformally
invariant \cite{takook} and in the present work, the related
two-point function was obtained with the following properties: It
is invariant under the conformal transformation and is free of
infrared divergence. The latter is achieved by carrying out the
calculation in Krein space. The result may be important on
formulation of the linear quantum gravity in de Sitter space.
Actually we believe that quantization in Krein space sheds some
light on the problem of the non-renormalizability of quantum
gravity.

\vspace{3mm} \noindent {\bf{Acknowledgments}}: We would like to
thank S. Ardeshirzade and E. Ariamand for their collaboration in
the early stage of this work. One of us, MRT, is grateful to S.
Fatemi for her useful comments. \setcounter{equation}{0}

\setcounter{equation}{0}
\begin{appendix}
\section{ SOME MATHEMATICAL PRELIMINARIES}

In this appendix we first review the Krein space briefly and then
collect some useful relations.\\Hilbert space is built by a set of
modes with positive norms:
$${\cal
H}= \bigg \{
\sum_{k\geq0}\alpha_k\phi_k;\sum_{k\geq0}|\alpha_k|^2<\infty\bigg\},
\,\,\mbox{with}\,\,(\phi_1,\phi_2)>0,$$ Krein space is defined as
a direct sum of a Hilbert space and an anti-Hilbert space
(negative inner product space):
$${\cal K}= {\cal
H}\oplus \bar {\cal H},$$ where $ \bar {\cal H}$ stands for the
anti-Hilbert space. Note that due to the indefinite inner product
space, some states are allowed to have negative norm. These modes
are only used as a mathematical tool in renormalization procedure
and are ruled out by imposing some conditions. In fact as
discussed in \cite{gazeaurenaud}, in Krein space setup, minimally
coupled scalar field is defined on non-Hilbertian Fock space. This
is followed by the fact that the one-particle sector is itself not
a Hilbert space since the total space (Krein space) is equipped
with an indefinite inner product. The physical space is the
quotient space: Krein space/negative-norm space. This is a
Hilbert space carrying the UIR of the de Sitter group. It is shown
that quantization in Krein space either removes some infinities
(for example the vacuum energy vanishes without any need of
reordering the terms), or at least regularizes the theory (for
more details see \cite{casimir} and references therein).

In what follows, some useful relations that are used in this paper, are listed:\\
$\bar\partial_{\alpha}$ is the tangential (or transverse)
derivative on dS space, defined by

$$\bar\partial_{\alpha}=\theta_{\alpha\beta}\partial^{\beta}=\partial_{\alpha}+x_{\alpha}x\cdot\partial,\,\,\,
\mbox{with}\,\,\,x\cdot\bar\partial=0\,,$$ and also one can define
\begin{equation}
\begin{aligned}
M_{\alpha\beta}\equiv&-i(x_{\alpha}\partial_{\beta}-x_{\beta}\partial_{\alpha})=
-i(x_{\alpha}\bar\partial_{\beta}-x_{\beta}\bar\partial_{\alpha}),\\
S_{\alpha\beta}\K_{\gamma\delta}\equiv&-i(\eta_{\alpha\gamma}\K_{\beta\delta}-\eta_{\beta\gamma}
\K_{\alpha\delta} +
 \eta_{\alpha\delta}\K_{\beta\gamma}-\eta_{\beta\delta}
 \K_{\alpha\gamma}).
 \end{aligned}
\end{equation}
Operator $Q^{(1)}_2$ commutes with the action of the group
generators, thus, it is constant in each UIR. The eigenvalues of
$Q^{(1)}_2$ can be used to classify the UIR's {\it i.e.,} \b
(Q^{(1)}_2-\langle Q^{(1)}_2\rangle){\cal K}(x)=0. \e Following
Dixmier \cite{dixmier}, one can get a classification scheme using
a pair $(p,q)$ of parameters involved in the following possible
spectral values of the Casimir operators:
\begin{equation}
Q^{(1)}_{p}=\left(-p(p+1)-(q+1)(q-2)\right)I_d ,\qquad\quad
Q^{(2)}_{p}=\left(-p(p+1)q(q-1)\right)I_d\,.
\end{equation}
Three types of UIR are distinguished for $SO(1,4)$ according to
the range of values of the parameters $q$ and $p$ \cite{dixmier,
takahashi}, namely: principal, complementary and discrete series.
The flat limit indicates that for the principal and complementary
series the value of $p$ bears the meaning of spin. For example in
discrete series $p=q=2$ have a Minkowskian interpretation as a
massless spin-2 particle.

The action of the Casimir operators $Q_1 $ and $ Q_2$ can be
brought in the more explicit form
\begin{equation}
\begin{aligned}
Q_1 K_{\alpha} =&(Q_0-2)K_{\alpha}+2x_{\alpha}\partial\cdot K
-2\partial_{\alpha} x\cdot K\\
=&(Q_0-2)K_{\alpha}+2x_{\alpha}\bar\partial\cdot K
-2\bar\partial_{\alpha} x\cdot K+2x_\alpha (x.K),
\end{aligned}
\end{equation}

\b Q_2 {\cal K}_{\alpha \beta}=(Q_0-6){\cal K}_{\alpha
\beta}+2{\cal S}x_{\alpha}\partial\cdot {\cal K}_{\beta}-2{\cal
S}\partial_{\alpha}x\cdot {\cal
K}_{\beta}+2\eta_{\alpha\beta}{\K}'. \e

The two-point function (\ref{twopoint}) with the choice of $x'$
reads
$$ {\cal W}_{\alpha\beta \alpha'\beta'}(x,x')=\Delta'_{\alpha\beta
\alpha'\beta'} (x,\partial,x',\partial'){\cal W}(x,x'),
$$ where
\begin{equation}
\begin{aligned}
\Delta'_{\alpha\beta
\alpha'\beta'}(x,\partial,x',\partial')=-\frac{2}{3}{\cal
S}\theta' \theta.\Big(\theta'.\theta
    -D'_{1}[x'.\theta+\theta.\bar\partial']\Big) +{\cal S}{\cal S}'\theta.\theta'\Big(\theta'.\theta
    -D'_{1}[x'.\theta+\theta.\bar\partial']\Big)\\
     +\frac{1}{3} D'_2{\cal
S}\Big(4(x'.{\theta}) + ({\theta}.{\bar\partial}') -
x'({\theta}.)\Big)\Big(\theta'.\theta
    -D'_{1}[x'.\theta+\theta.\bar\partial']\Big),
\end{aligned}
\end{equation}
note that, the primed operators act only on the primed
coordinates.

To obtain the two-point function, the following identities become
important
 \begin{equation} \bar{\partial}_\alpha
f({\cal{Z}})=-(x'.\theta_{\alpha})\frac{d
f(\cal{Z})}{d{\cal{Z}}},\end{equation}
\b\theta^{\alpha\beta}\theta'_{\alpha\beta}=\theta..\theta'=3+{\cal{Z}}^2,\;\;\
(x.\theta'_{\alpha'})(x.\theta'^{\alpha'})={\cal{Z}}^2-1,\;\;\
(x.\theta'_{\alpha})(x'.\theta^{\alpha})={\cal{Z}}(1-{\cal{Z}}^2),\e

\begin{equation}
\begin{aligned}
\bar{\partial}_\alpha(x.\theta'_{\beta'})&=\theta_{\alpha}.\theta'_{\beta'},\hspace{5mm}
\bar{\partial}_\alpha(x'.\theta_{\beta})=x_\beta(x'.\theta_{\alpha})-{\cal{Z}}\theta_{\alpha\beta},\\
\bar{\partial}_\alpha(\theta_{\beta}.\theta'_{\beta'})&=x_\beta(\theta_{\alpha}.\theta'_{\beta'})+
\theta_{\alpha\beta}(x.\theta'_{\beta'}),\hspace{3mm}
\theta'^{\beta}_{\alpha'}(x'.\theta_{\beta})=-{\cal{Z}}(x.\theta'_{\alpha'}),\\
\theta'^{\gamma}_{\alpha'}(\theta_{\gamma}.\theta'_{\beta'})&=\theta'_{\alpha'\beta'}+(x.\theta'_{\alpha'})(x.\theta'_{\beta'}),\\
 Q_0f({\cal{Z}})&=(1-{\cal{Z}}^2)\frac{d^2 f(\cal{Z})}{d{\cal{Z}}^2}-4{\cal{Z}}\frac{d
f(\cal{Z})}{d{\cal{Z}}},\\
{\int{\delta{({\cal{Z}})}d{\cal{Z}}}}&={\vartheta({\cal{Z}})},\;\;\
{\int{{\vartheta({\cal{Z}}}})}d{\cal{Z}}={\cal{Z}}{\vartheta({\cal{Z}})},\\
{\int{\frac{{\vartheta({\cal{Z}}-1})}{{\cal{Z}}+1}}}d{\cal{Z}}&={\vartheta({\cal{Z}}-1)}\Big({\ln{|{\cal{Z}}+1}|-\ln{2}}\Big),\\
 {\int {\vartheta({\cal{Z}}-1)}{\ln{|{\cal{Z}}+1}|}}
d{\cal{Z}}&={\vartheta({\cal{Z}}-1)}\Big({\ln{|{\cal{Z}}+1}|-1}\Big)({\cal{Z}}+1)-{\vartheta({\cal{Z}}-1)}(\ln4-2),\\
{\int{\delta'{({\cal{Z}})} f({\cal{Z}}) d{\cal{Z}}}} &=
-{\int{\delta{({\cal{Z}})} f'({\cal{Z}})
d{\cal{Z}}}}.\
\end{aligned}
\end{equation}

\setcounter{equation}{0}
\section{Mathematical relations underling the Eq. (\ref{3.9})}

Substituting $K$ in Eq. (\ref{ar}-II), yields
\begin{equation}\label{B.1}
Q_0^2(Q_0+2)\phi_3=(6Q_0^2+12Q_0-16)(x.Z_2)\phi_2+(12Q_0-16)(Z_2.\bar\partial)\phi_2,
\end{equation} then the general solution for
$\phi_3$ can be written as \b \label{B.2}\phi_3= c_1(x.Z_2)\phi_2
+ c_2(Z_2.\bar\partial)\phi_2, \e where $c_1$ and $c_2$ are two
constants. In order to find $c_1$ and $c_2$, we do the following
steps:\\ {\textbf{ Step(I)}}: the divergenceless condition
$(Q_0\phi_3=4x.Z_2\phi_2+Z_2.\bar\partial\phi_2)$, together with
(\ref{B.1}), results in \begin{equation}\label{B.3}
2(Q_0+4)(Q_0-2)(x.Z_2)\phi_2=(Q_0-2)(Q_0-8)(Z_2.\bar\partial)\phi_2,
\end{equation} on the other hand from Eq. (\ref{3.8}) we have \begin{equation}\label{B.4}
(Q_0-2)(Q_0+4)(Q_0-6)(x.Z_2)\phi_2=-6(Q_0-2)(Q_0-4)(Z_2.\bar\partial)\phi_2,
\end{equation} from (\ref{B.3}) and (\ref{B.4}), one obtains
\begin{equation}\label{B.5}
\begin{aligned}
Q_0(Q_0-2)^2(Z_2.\bar\partial)\phi _2&=0,\\
Q_0(Q_0-2)^2(Q_0+4)(x.Z_2)\phi_2&=0.
\end{aligned}
\end{equation} Substituting (\ref{B.5}) in (\ref{B.2}) results in
 \b\label{B.6} Q_0(Q_0-2)^2(Q_0+4)\phi_3=0.\e  After doing some straightforward algebra, one obtains the reduced forms of (\ref{B.5}) and (\ref{B.6}) as
follows:
 \b\label{b.11}
(Q_0-2)^2(Z_2.\bar\partial)\phi_2=0, \;\;\;\;\
(Q_0-2)^2(Q_0+4)(x.Z_2)\phi_2=0, \e or \b
(Q_0-2)^2(Q_0+4)\phi_3=0. \e {\textbf{ Step(II)}}: Using the
divergenceless condition and (\ref{B.1}), one gets
\begin{equation}\label{b.13}
Q_0(Q_0+4)(Q_0-2){\phi_3}=3(Q_0-4)(Q_0-2)(Z_2.{\bar\partial}){\phi_2}.\end{equation}
Combining (\ref{b.13}) and (\ref{B.4}) results in
\begin{equation}\label{b.15}
(Q_0+2)(Q_0+4)(Q_0-2)(x.Z_2)\phi_2=-2(Q_0+4)(Q_0-2)(Z_2.\bar\partial)\phi_2.
\end{equation} From (\ref{B.3}), (\ref{b.15}) and  (\ref{b.11}), we get \begin{equation}
Q_0(Q_0+4)(Q_0-2)(x.Z_2)\phi_2=-{\frac{1}{2}}Q_0(Q_0+4)(Q_0-2)(Z_2.\bar\partial)\phi_2
,\end{equation}  using the divergenceless condition and
(\ref{b.11}), the above equation can be written as \b\label{b.17}
Q_0^2(Q_0-2)(Q_0+4)\phi_3=-{\frac{1}{2}}Q_0^2(Q_0-2)(Q_0+4)(Z_2.\bar\partial)\phi_2.
\e From Eqs. (\ref{B.2}) and (\ref{b.17}), we obtain $
c_1-2c_2=1$, and substituting (\ref{B.2}) into (\ref{B.1})
together with (\ref{b.11}) results in $c_1=c_2=-1$.

\end{appendix}


\begin{thebibliography}{a}
\addcontentsline{toc}{chapter}{Bibliographie}

\bibitem{binegar} B. Binegar, C. Fronsdal and W. Heidenreich, Phys. Rev. D 27, (1983)
2249.

\bibitem{takook} M. V. Takook, M. R. Tanhayi and S. Fatemi,
J. Math. Phys. 51, (2010) 032503, arXiv:0903.5249v1.

\bibitem{3} This was first established for the Maxwell
equations in: H. Bateman, Proc. London Math. Soc. 7, (1909) 70; E.
Cunningham, Proc. London Math. Soc. 8, (1909) 77;\\ for massless
spin $1/2$ in: P.A.M. Dirac, Ann. Math. 37, (1936) 429;\\ and for
any spin in: A. McLennan, Nuovo Cim. 3, (1956) 1360; J. S. Lomont,
Nuovo Cim. 22, (1961) 673.

\bibitem{barut} A. O. Barut , A. B\"ohm, J. Math. Phys. 11, (1970) 2938.

\bibitem{angelo} E. Angelopoulos, M. Laoues, Rev. Math. Phys.
10, (1998) 271.

\bibitem{allen} B. Allen, M. Turyn, Nucl. Phys. B 292, (1987) 813;.

\bibitem{floratos} E. G. Floratos, J. Iliopoulos, T. N. Tomaras, Phys. Lett. B
197, (1987) 373.

\bibitem{antoniadis} I. Antoniadis, E. Mottola, J. Math. Phys. 32, (1991) 1037.

\bibitem{ford} H. L. Ford, Phys. Rev. D 31, (1985) 710.

\bibitem{antoni1} I. Antoniadis, J. Iliopoulos, T. N. Tomaras, Phys. Rev. Lett. 56, (1986) 1319.

\bibitem{tsamis} N. C. Tsamis, R. P. Woodard, Phys. Lett. B 292, (1992)
269; Commun. Math. Phys. 162, (1994) 217.

\bibitem{antoni} I. Antoniadis, J. Iliopoulos, and T. N. Tomaras, Nucl. Phys. B 462, (1996) 437.

\bibitem{higuchi}  A. Higuchi, S. S. Kouris, Class. Quant. Grav.
17, (2000) 3077; \emph{ibid}, Class. Quant. Grav. 20, (2003) 3005.

\bibitem{vega} H. J. de Vega, J. Ramirez, and N. Sanchez, Phys. Rev.
D 60, (1999) 044007; S. W. Hawking, T. Hertog, and N. Turok, Phys.
Rev. D 62, (2000) 063502.

\bibitem{takook2} M. V. Takook, Proceeding of the Wigsym6, August, 1999,
Istanbul, Turkey; M. V. Takook, IPJ., 3-1 (2009) 1-8.

\bibitem{dehghani} M. Dehghani, S. Rouhani, M. V. Takook and M. R.
Tanhayi, Phys. Rev. D 77, (2008) 064028.

\bibitem{allenfolacci} B. Allen, B. Folacci, Phys. Rev. D 35, (1987)
3771.

\bibitem{gazeaurenaud} J. P. Gazeau, J. Renaud, and M. V. Takook, Class
Quant. Grav. 17, (2000) 1415.

\bibitem{riess} A. G. Riess et al. [Supernova Search Team Collaboration], Astro. J. 116,  (1998) 1009; S. Perlmutter et al. [Supernova Cosmology Project
Collaboration], Astro. J. 517,  (1999) 567; U. Seljak, A. Slosar,
and P. McDonald, JCAP 014, (2006) 610; A. G. Riess et al., Astro.
J. 98, (2007) 659.

\bibitem{linde} A. D. Linde, Harwood Academic Publishers, Chur,
Switzerland. (1990) {\it PARTICLE PHYSICS AND INFLATIONARY
COSMOLOGY}.

\bibitem{perlmutter} P. Perlmutter, et. al., Astro. J. 517, (1999) 565;
A. Jaros and M. E. Peskin Int. J. Mod. Phys. A 715, (2000) 1581.

\bibitem{gazeauhans} J. P. Gazeau, M. Hans, J. Math Phys. 29, (1988) 2533.

\bibitem{levy} M. Levy-Nahas, J. Math. Phys. 8, (1967) 1211.

\bibitem{bacry} H. Bacry, J. M. Levy-Leblond, J. Math. Phys. 9, (1968) 1605.

\bibitem{dirac} P. A. M. Dirac, Ann. of Math. 36, (1935) 657.


\bibitem{mack} G. Mack and A. Salam, Ann. Phys. 53, (1969) 174.

\bibitem{kastrup} H. A. Kastrup, Phys. Rev. 150, (1964) 1189; C. R. Preitschop,
M. A. Vosiliev, Nucl. Phys. B 549, (1999) 450.

\bibitem{behroozi} S. Behroozi, S. Rouhani, M. V. Takook and M. R. Tanhayi,
Phys. Rev. D 74, (2006) 124014.

\bibitem{garidigatak} T. Garidi, J. P. Gazeau and M. V. Takook, J. Math.
Phys. 44, (2003) 3838.

\bibitem{garidigarou} T. Garidi, J. P. Gazeau, S. Rouhani, M. V. Takook, J. Math. Phys. 49, (2008) 032501, gr-qc/0608004.

\bibitem{gazeautakook} J. P. Gazeau, M. V. Takook, J. Math. Phys. 41, (2000)
5920.

\bibitem{bros} J. Bros, J. P. Gazeau, and U. Moschella, Phys. Rev. Lett.
73, 1746 (1994).

\bibitem{bros2} J. Bros and U. Moschella, Rev. Math. Phys. 8, 327
(1996).

\bibitem{allen2} B. Allen, T. Jacobson, Comm. Math. Phys. 103, (1986)
669.

\bibitem{folacci} A. Folacci, J. Math. Phys. 32, (1991) 2828.

\bibitem{takook3} M. V. Takook, Mod. Phys. Lett. A 16, (2001) 1691.

\bibitem{chernikov} N. A. Chernikov and E. A. Tagirov, Ann. Inst. Henri
Poincar\'{e}, IX (1968) 109.

\bibitem{JHEP} M. V. Takook and M. R. Tanhayi, JHEP 1012, (2010)
044.

\bibitem{stelle} see for example, K. S. Stelle, Phys. Rev. D 16, (1977) 953; E. S.
Fradkin and A. A. Tseytlin, Nucl. Phys. B 201, (1982) 469.

\bibitem{bender} C. M. Bender and P. D. Mannheim,
Phys. Rev. Lett. 100, (2008) 110402; ibid, Phys. Rev. D 78, (2008)
025022; C. M. Bender and S. Boettcher, Phys. Rev. Lett. 80, (1998)
5243.

\bibitem{Rodrigues} D. C. Rodrigues Phys. Rev. D 77, (2008) 023534.

\bibitem{casimir} M. V. Takook, H. Pejhan, M. Tanhayi-Ahari and M. Reza
Tanhayi, \emph{Casimir Effect For a Scalar Field via Krein
Quantization}, arXiv: 1204.6001.

\bibitem{dixmier} J. Dixmier, Bull. Soc. Math. France 89, (1961) 9.

\bibitem{takahashi} B. Takahashi, Bull. Soc. Math. France 91, (1963) 289.


\end{thebibliography}
\end{document}